\documentclass[twocolumn,prx,superscriptaddress]{revtex4-1}
\usepackage{amsmath}
\usepackage{bm}
\usepackage{graphicx}
\usepackage{epstopdf}
\usepackage{color}
\usepackage{float}

\begin{document}
\title{Pinch points and half-moons in dipolar-octupolar Nd$_2$Hf$_2$O$_7$   }

\author{A. Samartzis}
\altaffiliation{alexandros.samartzis@helmholtz-berlin.de}
\affiliation{\mbox{Helmholtz-Zentrum Berlin f\"ur Materialien und Energie, Hahn-Meitner-Platz 1, 14109 Berlin, Germany}}
\affiliation{\mbox{Institut f\"ur Festk\"orperphysik, Technische Universit\"at Berlin, Hardenbergstr. 36, 10623 Berlin, Germany}}

\author{J. Xu}
\affiliation{\mbox{Helmholtz-Zentrum Berlin f\"ur Materialien und Energie, Hahn-Meitner-Platz 1, 14109 Berlin, Germany}}
\affiliation{\mbox{Heinz Maier-Leibnitz Zentrum at Lichtenbergstr. 1, 85748, Garching, Germany}}

\author{V. K. Anand}
\affiliation{\mbox{Helmholtz-Zentrum Berlin f\"ur Materialien und Energie, Hahn-Meitner-Platz 1, 14109 Berlin, Germany}}
\affiliation{\mbox{Department of Physics, University of Petroleum and Energy Studies, Dehradun, Uttarakhand, 248007, India}}

\author{A. T. M. N. Islam}
\affiliation{\mbox{Helmholtz-Zentrum Berlin f\"ur Materialien und Energie, Hahn-Meitner-Platz 1, 14109 Berlin, Germany}}

\author{J. Ollivier}
\affiliation{\mbox{Institut Laue Langevin, 6 rue Jules Horowitz,BP 156, F-38042 Grenoble, France}}

\author{Y. Su}
\affiliation{\mbox{J\"ulich Centre for Neutron Science at Heinz Maier-Leibnitz Zentrum, Forschungszentrum J\"ulich, Garching, Germany}}

\author{B. Lake}
\altaffiliation{bella.lake@helmholtz-berlin.de}
\affiliation{\mbox{Helmholtz-Zentrum Berlin f\"ur Materialien und Energie, Hahn-Meitner-Platz 1, 14109 Berlin, Germany}}
\affiliation{\mbox{Institut f\"ur Festk\"orperphysik, Technische Universit\"at Berlin, Hardenbergstr. 36, 10623 Berlin, Germany}}
\date{\today}

\begin{abstract}

While it is established that the pinch point scattering pattern in spin ice arises from an emergent coulomb phase associated with magnetic moment that is divergence-free, more complex Hamiltonians can introduce a divergence-full part. If these two parts remain decoupled, they give rise to the co-existence of distinct features. Here we show that the moment in ${\rm Nd_2Hf_2O_7}$ forms a static long-range ordered ground state, a flat, gapped pinch point excitation and dispersive excitations. These results confirm recent theories which predict that the dispersive modes, which arise from the divergence-full moment, host a pinch point pattern of their own, observed experimentally as `half-moons'.

\end{abstract}

\maketitle

The observation of the ‘pinch point pattern’ in magnetic materials signals the presence of an emergent coulomb phase where the divergence of the emergent vector field (in this case the magnetization) is zero (divergence-free) \cite{Henley}. The pinch point pattern arises in classical spin-ice materials such as ${\rm Ho_2Ti_2O_7}$ \cite{Bramwell2001,Fennell2009} or ${\rm Dy_2Ti_2O_7}$ \cite{morris2009} where the magnetic rare earth ions which lie on the verticies of a network of corner-share tetrahedra (pyrochlore lattice) are coupled by the ferromagnetic (FM) interactions and have strong local Ising anisotropy that forces the spins to point into or out of the center of every tetrahedron. This highly frustrated combination of FM interactions and Ising anisotropy gives rise to the ice rule that states that in the ground state 2 spins point into and 2 spins point out of each tetrahedron (2I2O configuration) leading to a highly degenerate ground state manifold \cite{Hertog2000}. Excitations from this state (such as 3 spins pointing into and 1 spin pointing out of a tetrahedron) are magnetic monopoles \cite{morris2009,Fennell2009,castelnovo2008}. Quantum spin ice can arise in materials where additional weak terms in the Hamiltonian allow mixing of the ground state manifold to give a spin liquid state \cite{Hermele2004,Shannon2012}. Violations of the ice-rule and the divergence-free condition allow monopole creation in the ground state, observed experimentally as a broadening of the pinch points \cite{Benton2012}.

\begin{figure}
	\centering
	\begin{tabular}{c @{\qquad} c }
		\includegraphics[width=1\columnwidth,keepaspectratio]{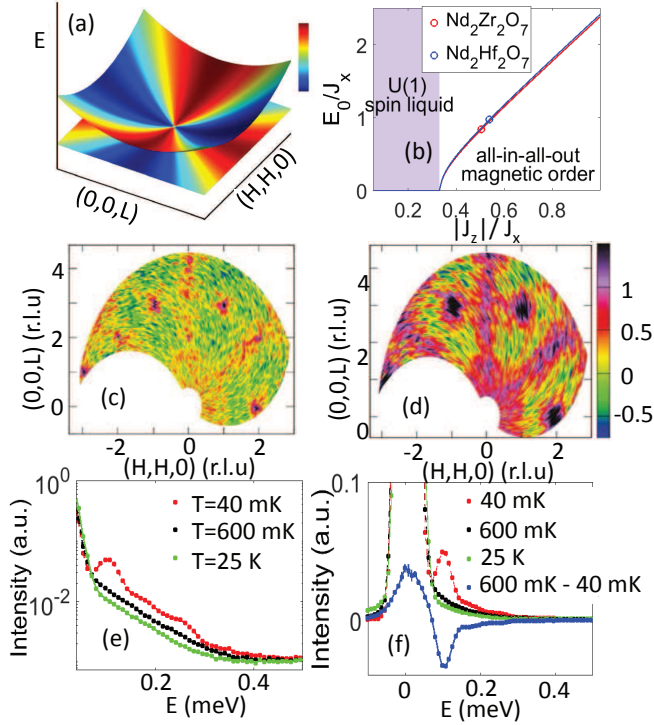} 
	\end{tabular}
	\caption{(a) Theoretical spectrum as a function of wavevector in the $[H,H,L]$-plane and energy, showing the flat pinch point mode and the dispersive half-moon mode which touch at the pinch points, the colours indicate the intensity of the dynamic structure factor \cite{HanYan2018}. (b) Phase diagram of the XYZ Hamiltonian showing the gap of the pinch point mode, $E_0=\sqrt{(3| \widetilde{J}_{\widetilde{z}}|- \widetilde{J}_{\widetilde{x}})(3| \widetilde{J}_{\widetilde{z}}|- \widetilde{J}_{\widetilde{y}})}$ as a function of $|\widetilde{J}_{\widetilde{z}}|/\widetilde{J}_{\widetilde{x}}$ \cite{Benton2016}. The circles locate ${\rm Nd_2Hf_2O_7}$ and ${\rm Nd_2Zr_2O_7}$. (c-d) $[H,H,L]$ wavevector map of the background-subtracted Z-Spin-flip PND measured at (c) 88~mK and (d) 600~mK. The presence of signal at the nuclear Bragg peaks, (2,2,2) and (0,0,2) is due to the subtraction of two large numbers. (e) INS spectra at $T=40$~mK, $T=600$~mK and $T=25$~K integrated over a large wavevector range ($[H,H,L]$: $-1<H<1.5$; $0.5<L<2.5$) and plotted on a log-scale as a function of energy, with no background subtraction or normalization. (f) The same data on a linear scale including a plot of the difference between the 600~mK and 40~mK datasets, the Bragg peaks were removed before the subtraction.}
	\label{fig:overview}
\end{figure}

\begin{figure}
	\centering 
	\includegraphics[width=1\columnwidth,keepaspectratio]{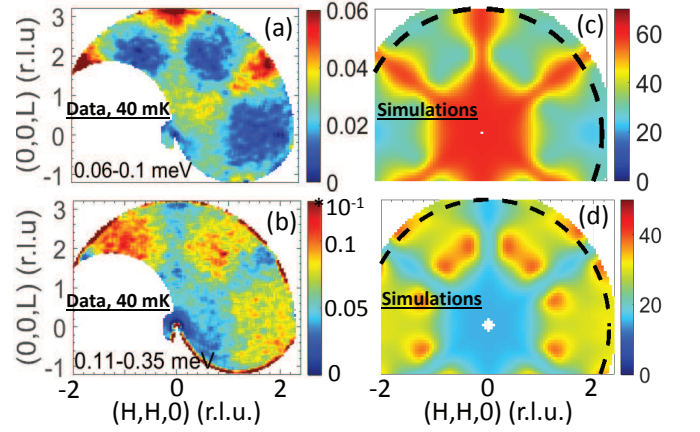}
	\caption{Comparison of the experimental and theoretical spectrum of ${\rm Nd_2Hf_2O_7}$ in the $[H,H,L]$ plane. (a-b) show the INS data at 40~mK integrated over the flat ($0.06 < E < 0.1$~meV) and dispersive ($0.11 < E < 0.35$~meV) modes respectively. The data were also integrated over the out-of-plane wavevector $[K,-K,0]$ by ${K= \pm 0.1}$~r.l.u\@ and they were normalized and the background was subtracted \cite{SM}. (c-d) show the corresponding neutron scattering structure factor simulated for the Hamiltonian $H_{XYZ}^{\rm DO}$ (Eq.~\eqref{eq:Hamilt}) using linear spin-wave theory \cite{SM} integrated over the same energy and wavevector ranges.}
	\label{fig:INS2}
\end{figure} 

More complex Hamiltonians can give rise to further changes in the magnetic properties. A particularly interesting phenomenon proposed recently is ‘magnetic moment fragmentation’ where the magnetic Hamiltonian contains both divergence-free and divergence-full parts \cite{MomFrag}. These sectors can remain decoupled, allowing for example, for the co-existence of long-range magnetic order in the form of a monopole crystal (divergence-full) alongside a static pinch point pattern (divergence-free). 
Related phenomena were observed in ${\rm Nd_2Zr_2O_7}$. 
Here the dipolar-octupolar doublet of the ${\rm Nd^{3+}}$ ion \cite{Xu2015, Lhotel2015}, gives rise to a Hamiltonian that has both FM and antiferromagnetic (AFM) interactions for different spin components as well as Ising anisotropy. The weak unfrustrated AFM term gives rise to long-range magnetic order with the all-in/all-out (AIAO) configuration where all the spins point into or all out of each tetrahedra (divergence-full) \cite{Lhotel2015, Xu2016}. In contrast the FM term gives rise to a pinch point pattern which is dynamic forms a divergence-free excitation \cite{Petit2016, Xu2019}. Application of a magnetic field was found to induce further exotic behaviors \cite{Lhotel2018,Xu2018,Opherden2017}.

Here we investigate another feature related to moment fragmentation, namely `half-moons'. These are rings observed in inelastic neutron scattering where the intensity is strongly modulated around the ring, rather than the uniform intensity distribution typical of conventional magnetic excitations such as spin-waves. Whilst they have occasionally been glimpsed in experiments \cite{Guitteny2013,Fennell2014, Petit2016,Lhotel2018} as well as in numerical simulations \cite{Taillefumier2014,Robert2008,Rau2016,Udagawa2016,Mizoguchi2017,Saha2021}, they were not studied systematically and their significance was not explained. Very recently the half-moons were investigated by two theoretical groups who showed that they arise from divergence-full moment and signal proximity to an emergent coulomb phase \cite{Mizoguchi2018, HanYan2018}.  

In this paper we study the pyrochlore ${\rm Nd_2Hf_2O_7}$  using neutron scattering measurements and show that it hosts half-moons. Previous investigations of ${\rm Nd_2Hf_2O_7}$ reveal that the crystal electric field surrounding the magnetic ${\rm Nd^{3+}}$ ion results in a Kramers doublet ground state well-isolated from the first excited state \cite{Anand2017}. The ground state wavefunction of the ${\rm Nd^{3+}}$ ion is of dipolar-octupolar nature with Ising anisotropy and the $g$-tensor was found to be $g_{zz}\approx 5.1$, $g_{xx}=g_{yy}=0$ within the pseudo-spin-1/2 model where $z$ is the local [1,1,1]-axis pointing towards the center of each tetrahedron \cite{Anand2017}. ${\rm Nd_2Hf_2O_7}$, orders below $T_{\rm N}=0.55$~K in the AIAO magnetic structure \cite{Anand2015} indicating the presence of an AFM interaction. The ordered moment at $T=0.1$~K is $m=0.62$~$\mu_B$ \cite{Anand2015} which is strongly reduced compared to the effective moment of the ground state wavefunction of 2.45~$\mu_B$  indicating the presence of quantum fluctuations as also inferred from muon spin relaxation measurements \cite{Anand2017}. ${\rm Nd_2Hf_2O_7}$  was also found to demonstrate an inverted magnetic hysteresis loop for magnetic fields along the [1,1,1] and [0,0,1] directions \cite{Opherden2018}.

In this present article we find that AIAO antiferromagnetic long-range order of ${\rm Nd_2Hf_2O_7}$ co-exists with a dynamical pinch point pattern that takes the form of a flat gapped mode. In addition there are higher energy dispersive modes which give rise to half-moon features. We perform a detailed examination of these dispersive modes and the associated half-moons and achieve excellent agreement between our data and simulations. Our results confirm the recent theories \cite{Mizoguchi2018, HanYan2018} that predict that these modes, which arise from the divergence-full part of the moment, also host a pinch point pattern of their own. 

\begin{figure}
	\includegraphics[width=1\columnwidth,keepaspectratio]{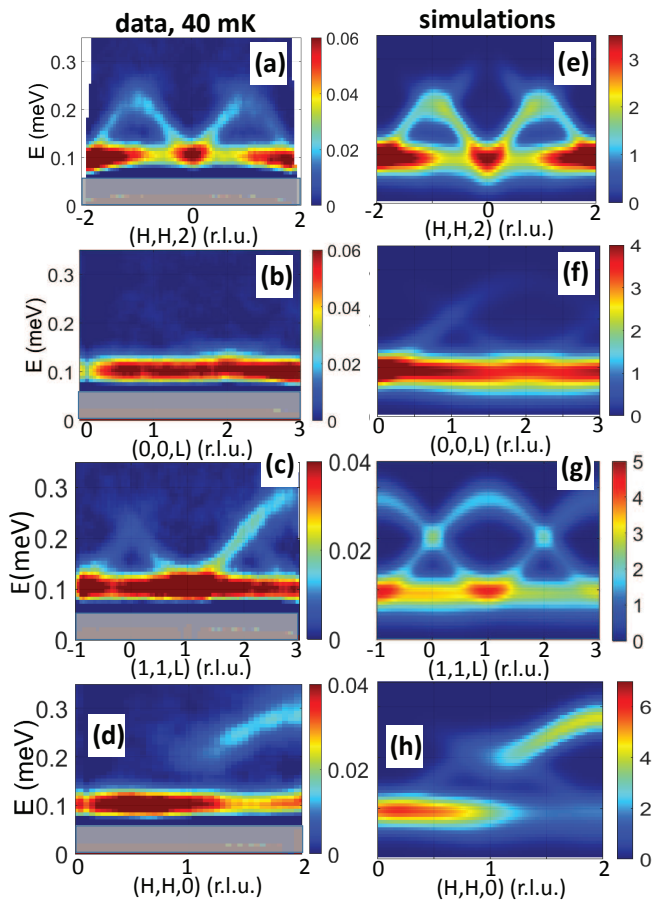}
	\caption{Energy-wavevector maps comparing the experimental and theoretical spectrum of ${\rm Nd_2Hf_2O_7}$. (a-d) show the INS measurement at 40~mK along various reciprocal space directions. The data were integrated over the out-of-plane and the in-plane wavevectors perpendicular to the slice by ${\rm \pm 0.2}$ r.l.u\@, and were normalized and background-subtracted \cite{SM}. The data below 0.06~meV (shaded region) are unreliable due to background oversubtraction. (e-h) show the corresponding neutron scattering structure factor simulated for $H_{XYZ}^{\rm DO}$ (Eq.~\eqref{eq:Hamilt}) using linear spin-wave theory \cite{SM} for the same wavevector integrations.}
	\label{fig:INS1}
\end{figure}
	
Neutron scattering was performed on a large single crystal of ${\rm Nd_2Hf_2O_7}$ of mass $\approx6$~g oriented to access the [$H,H,L$] reciprocal lattice plane. Inelastic neutron scattering (INS) was measured at the Institute Laue Langevin, France, on the direct geometry, time-of-flight spectrometer IN5 \cite{IN5}, with neutrons of wavelength 6~\AA\ and energy resolution 0.041~meV. $\omega$-scans were measured far below $T_{\rm N}=0.55$~K at 40~mK, just above $T_{\rm N}$ at 600~mK as well as at 25~K which was used to estimate the non-magnetic background. The data were background subtracted, normalized and analyzed using the Horace software \cite{Horace} (see Ref.\ \cite{SM}). Polarized neutron diffraction (PND) was measured using DNS at the Maier-Leibniz Zentrum, Germany \cite{DNS} with neutrons of wavelength 4.2~\AA. The neutrons were polarized along the (vertical) $Z$-direction and both spin-flip ($Z$SF) and non-spin-flip ($Z$NSF) channels were measured for $\omega$-scans  at 88~mK, 600~mK and 25~K.    

Figure~\ref{fig:overview}(c) presents the PND data below $T_{\rm N}$ at 88~mK. The data show a pinch point pattern co-existing with the magnetic Bragg peaks of the AIAO order at (2,2,0), (1,1,3) and (-1,-1,3). The pinch points, expected at (0,0,2), (1,1,1) and (2,2,2), are not sharp revealing that the ice-rule is not perfectly obeyed due to the presence of additional interactions beside the Ising term between the Nd$^{3+}$ ions. The data gives the magnetic signal integrated over energy, to resolve the energy scales of the various features we turn to the INS data. Fig.~\ref{fig:overview}(e) shows a cut through the INS data plotted as a function of energy. Below  $T_{\rm N}$ at 40~mK, a peak is observed at $E\approx0.1$~meV. A second broad peak occurs at $\approx 0.28$~meV, while by 0.4~meV the scattering approaches the non-magnetic background provided by the 25~K data. 

Wavevector maps in the $[H,H,L]$-plane of the INS data at 40~mK are shown in Fig.~\ref{fig:INS2}. For $0.06<E<0.1$~meV (Fig.~\ref{fig:INS2}(a)) which corresponds to the first peak in Fig.~\ref{fig:overview}(e), a clear pinch point pattern is found where the pinch points appear sharper than those observed in the PND data (Fig.~\ref{fig:overview}(c)). At energies below this peak, the pinch point pattern has disappeared. These results show that the pinch point pattern is from a dynamic gapped mode well separated in energy from the AIAO ground state. Energy versus wavevector slices through the INS data at $T=40$~mK are displayed in Fig.~\ref{fig:INS1} for several directions. The most striking feature is the sharp, gapped, dispersionless mode lying at an energy of $E_0 = 0.094~{\rm meV}$. This is the pinch point mode observed in the constant energy cut at $0.06<E<0.1$~meV (Fig.~\ref{fig:INS2}(b)). At higher energies there are sharp dispersive modes. 

These measurements reveal the unusual coexistence, in ${\rm Nd_2Hf_2O_7}$, of a long-range AIAO magnetically ordered ground state with a dynamical pinch point pattern associated with the 2I2O spin fluctuations of an emergent Coulomb phase. The divergence-full and divergence-free parts of the magnetic moment are responsible for these two phenomena respectively.
%The pinch point pattern is associated with the 2I2O spin configuration of the emergent Coulomb phase, while the magnetic Bragg peaks are associated with the AIAO spin configuration. The unusual coexistence of these two features suggest the phenomenon of moment fragmentation first predicted by M.\ E.\ Brooks-Bartlett {\it et al} \cite{MomFrag} where the magnetic moment fragments into divergence-free and divergence-full parts that are responsible for these two phenomena respectively. 
Similar magnetic behavior was observed in the related compound ${\rm Nd_2Zr_2O_7}$ \cite{Petit2016,Xu2019}.
 
Here we use the XYZ Hamiltonian, proposed for ${\rm Nd_2Zr_2O_7}$ \cite{Benton2016, Huang2014} to describe ${\rm Nd_2Hf_2O_7}$ and obtain the interactions:
\begin{equation}
\label{eq:Hamilt}
H_{XYZ}^{DO}=\sum_{<ij>} \widetilde{J}_{\widetilde{x}}\widetilde{\tau}_i^{\widetilde{x}}\widetilde{\tau}_j^{\widetilde{x}}+\widetilde{J}_{\widetilde{y}}\widetilde{\tau}_i^{\widetilde{y}}\widetilde{\tau}_j^{\widetilde{y}}+
\widetilde{J}_{\widetilde{z}}\widetilde{\tau}_i^{\widetilde{z}}\widetilde{\tau}_j^{\widetilde{z}}
\end{equation} 
where $\widetilde{\tau}_{i}^{\widetilde{\alpha}}$ is the $\widetilde{\alpha}$-component of the $i^{th}$ pseudospin-1/2 operator and $\widetilde{J}_{{\widetilde{\alpha}}}$ is the exchange interaction between the $\widetilde{\alpha}$ components of nearest neighbor spins. A rotated local coordinate system is used where $\widetilde{z}$ is rotated by an angle $\theta$ from the local, (1,1,1)-axis \cite{SM,Ross2011}. 

The Hamiltonian was determined by fitting linear spin-wave theory \cite{TothPRX} to the INS data \cite{SM}. The results presented on the left hand side of Fig.~\ref{fig:INS2} and Fig.~\ref{fig:INS1} agree excellently with the data. The fits yielded: $\widetilde{J}_{\widetilde{x}}=0.106(5)$~meV, $\widetilde{J}_{\widetilde{y}}=0.008(5)$~meV, $\widetilde{J}_{\widetilde{z}}=-0.057(4)$~meV, while $\theta\approx72.5^{\circ}$ was obtained from the Curie-Weiss temperature \cite{SM}. $\widetilde{J}_{\widetilde{x}}$ is ferromagnetic and would, on its own, give rise to classical, highly-degenerate, `rotated spin-ice' ground state with a 2I2O spin configuration characterized by a pinch point pattern at zero energy transfer. The addition of a weak antiferromagnetic $\widetilde{J}_{\widetilde{z}}$ introduces quantum fluctuations which mix the extensive ground state manifold to give a $U$(1) spin liquid \cite{Benton2016}. For $|\widetilde{J}_{\widetilde{z}}|>\widetilde{J}_{\widetilde{x}}/3$, a long-range AIAO ordered ground state is induced, and the pinch point becomes dynamic and gapped as is the case for ${\rm Nd_2Hf_2O_7}$. Fig.~\ref{fig:overview}(b) shows the phase diagram with the gap plotted as a function of $|\widetilde{J}_{\widetilde{z}}|$ and the location of ${\rm Nd_2Hf_2O_7}$ proximate to the $U(1)$ spin liquid. 

\begin{figure}
	\centering
	\begin{tabular}{c @{\qquad} c }
		%\hspace*{-1.5cm}
		%\vspace*{-0.2cm}
		\includegraphics[width=1\columnwidth,keepaspectratio]{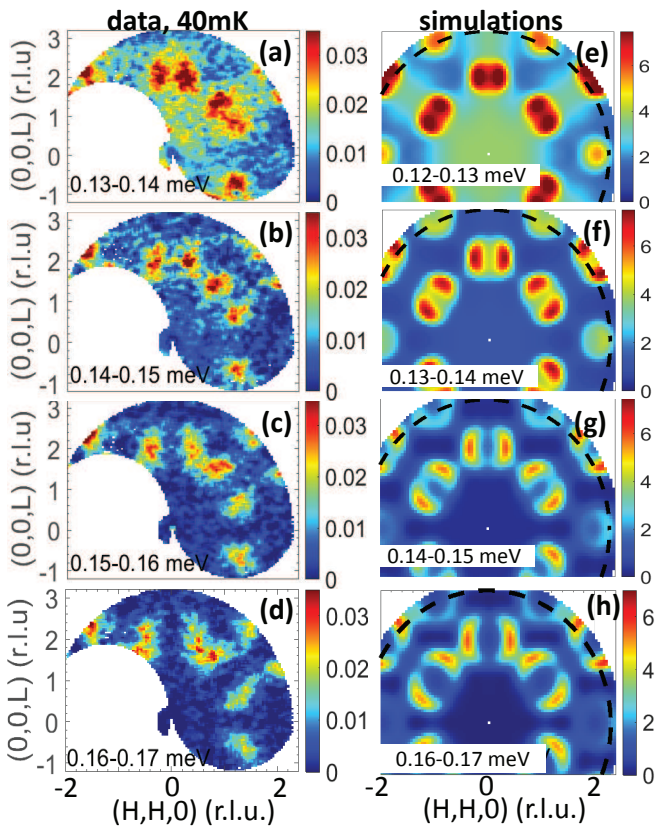} 
	\end{tabular}
	\caption{Wavevector maps at constant energy comparing the experimental and theoretical spectrum. (a-d) show the INS measurement at 40~mK in the $[H,H,L]$-plane integrated over the specified energy ranges. The data were normalized, background-subtracted, and integrated over the out-of-plane wavevector by ${\pm 0.15}$~r.l.u\@.  (e-h) show the corresponding neutron scattering structure factor simulated for $H_{XYZ}^{\rm DO}$ (Eq.~\eqref{eq:Hamilt}) using linear spin-wave theory \cite{SM} for the same wavevector integration. For better correspondence with the experiment, the energy integration is systematically shifted downwards by 0.01~meV which is within the limit of the errorbars of the exchange interactions.}
	\label{fig:halfMoon}
\end{figure}

The origins of this apparent magnetic moment fragmentation for the XYZ Hamiltonian were investigated by O. Benton and found to be a consequence of the dipolar-octupolar nature of the Kramers doublet ground state of Nd$^{3+}$ and the equations of motion of the magnetic degrees of freedom \cite{Benton2016}. He proposed that the AIAO magnetic order (due to $\widetilde{\tau}_{i}^{\widetilde{z}}$) is entirely divergence-full, while the dynamics (due to $\widetilde{\tau}_{i}^{\widetilde{x}}$ and  $\widetilde{\tau}_{i}^{\widetilde{y}}$) completely decouple into a divergence-free part responsible for the flat pinch point mode and a divergence-full part which gives rise to the dispersive modes at higher energies. 

Because of its proximity to the $U(1)$ quantum spin liquid state we would expect ${\rm Nd_2Hf_2O_7}$ to show signatures of this spin liquid just above $T_{\rm N}\approx 0.55$~K when the AIAO magnetic order is lost. Fig.~\ref{fig:overview}(d) shows the PND measurement at $T=0.6$~K ($>T_{\rm N}$), here the pinch point pattern is enhanced compared to $T=0.088$~K (Fig.~\ref{fig:overview}(c)) and the pinch points are broadened. In addition the AIAO Bragg peaks at $(2,2,0)$, $(3,1,1)$ and $(3,-1,-1)$ have been replaced by broad features. Fig.~\ref{fig:overview}(e) shows an energy cut through the INS data at 0.6~K. The pinch point mode at 0.094~meV as well as the 0.28~meV peak found at $T=0.04$~K, have been replaced by broad magnetic signal peaked at the elastic energy. This is shown clearly in Fig.~\ref{fig:overview}(f) where the difference is taken between the cuts at 0.6~K and 0.04~K, revealing that the peak found at 0.094~meV below $T_{\rm N}$ is replaced by a peak at $\approx0$~meV above $T_{\rm N}$. Similar results were observed for ${\rm Nd_2Zr_2O_7}$ where above $T_{\rm N}$ the pinch point was found at elastic energies. Here the transition was attributed to a Higgs transition where the emergent gauge field of the Coulomb phase above $T_{\rm N}$ becomes gapped by the condensation of emergent gauge charges or monopoles into the ground state below $T_{\rm N}$ \cite{Xu2020}. Another study however found more complex behavior with co-existence of inelastic and elastic pinch points patterns above $T_{\rm N}$ \cite{Leger2021}. 

We now return to low temperatures and take a closer look at the high energy dispersive excitations ($E>0.11$~meV) which show several unconventional features. Firstly these modes have maximum energy at the wavevectors of the magnetic Bragg peaks ((1,1,2) and (2,2,0)) rather than being minimum there as is usually the case for an ordered system (Fig.~\ref{fig:INS2}-\ref{fig:INS1}). Instead their energy minima occur at the wavevectors of the pinch points (0,0,2), (1,1,1) and (2,2,2) where they touch the flat gapped mode. Furthermore, an integration of the dispersive modes over energy ($0.11<E<0.35$~meV) shows that they have maximum intensity close to these pinch point wavevectors, but curiously the intensity distribution is opposite to that of the flat pinch point mode (Fig.~\ref{fig:INS2}). The flat mode shows the traditional pinch point pattern with a streak of intensity along radial wavevectors (Fig.~\ref{fig:INS2}(a)), but the dispersive modes have the opposite intensity distribution with signal concentrated at transverse wavevectors (Fig.~\ref{fig:INS2}(b)). Finally wavevector maps for narrow energy integration ranges just above the minima of the dispersive modes, do not show the rings typically observed when slicing through cone-shaped dispersions, instead `split-ring' or `half-moon' features surrounding the pinch point wavevectors are observed with opposite intensity distribution to the flat mode (Fig.~\ref{fig:halfMoon}). By taking a series of energy slices we show that the half-moons disperse away from the pinch point with increasing energy but never become complete rings. Their behavior is well-captured by the spin-wave simulations which show good agreement with the data in Fig.~\ref{fig:halfMoon}.

Half-moons have been glimpsed occasionally in INS data from pyrochlore magnets such as ${\rm Tb_2Ti_2O_7}$ \cite{Guitteny2013,Fennell2014} and ${\rm Nd_2Zr_2O_7}$ \cite{Petit2016,Lhotel2018} as well as in numerical simulations of kagome- and pyrochlore-related magnets \cite{Taillefumier2014,Robert2008,Rau2016,Udagawa2016,Mizoguchi2017,Saha2021}, however a systematic experimental investigation has not been performed, nor has an explanation for their existence been given until recently. Mizoguchi {\it et al} \cite{Mizoguchi2018} showed that half-moons can be interpreted as shadows of pinch points, signaling proximity to a Coulomb phase. From a real-space perspective, they reflect the formation of magnetic clusters involving short-range correlation of conserved spin e.g.\ due to further neighbor interactions in Heisenberg magnets. H. Yan {\it et al} \cite{HanYan2018} used a Helmholtz decomposition to make the connection between traditional pinch points (satisfying the divergence-free condition) and half-moons (satisfying the divergence-full condition) proving they originate from the same, proximate, $U$(1) gauge symmetry. They further showed that half-moons themselves arise from a rotated pinch point pattern that is oriented perpendicular to the conventional pinch point pattern and exists on a cone-shaped dispersing mode centered at the pinch point wavevector. A schematic of the pinch point and half-moon modes which illustrates their energies and intensity distribution is given in Fig.~\ref{fig:overview}(a). 

The half-moon mode in ${\rm Nd_2Hf_2O_7}$ can be seen in Fig.~\ref{fig:INS1} dispersing out of (0,0,2) which has high intensity along the transverse $(H,H,2)$ direction (Fig.~\ref{fig:INS1}(a)), but no intensity along the longitudinal  $(0,0,L)$ direction of the traditional pinch point (Fig.~\ref{fig:INS1}(b)). The half-moons shown in Fig.~\ref{fig:halfMoon}, which are observed when taking constant-energy slices through this dispersing mode, are due to its highly directional intensity distribution. Integrating over the dispersing half-moon mode reveals its rotated pinch point pattern with opposite intensity distribution to that of the flat pinch point mode (compare Fig.~\ref{fig:INS2}(a) and (b)). 

To conclude, we have investigated the magnetic properties of ${\rm Nd_2Hf_2O_7}$ and found it to 
separate into distinct 
%demonstrate clear evidence of magnetic moment fragmentation into 
uncoupled divergence-free and divergence-full parts which give rise to separate features. Long-range AIAO spin order (divergence-full) co-exists with a dynamic coulomb phase characterized by 2I2O fluctuations observed as a flat, gapped, pinch point mode (divergence-free).  Additional dispersive excitations (divergence-full) are observed at high energies which touch the flat mode at the pinch points. We have performed a detailed experimental investigation along with spin-wave calculations of the high energy excitations. Our results confirm recent theories that predict that these modes also host a pinch point pattern perpendicular to the conventional pinch point direction with the opposite intensity distribution \cite{Mizoguchi2018,HanYan2018}. Because of the dispersing nature of this pinch point pattern, it is observed as half-moons in constant energy slices through the INS data, very different from the uniform rings observed for conventional spin-waves. Half-moons therefore signal proximity to a coulomb phase where violation of spin-ice rules allow a divergence-full part to exist that forms a dispersive pinch point pattern in its own right corresponding to monopole propagation.

\begin{acknowledgments}
	B.L. acknowledges the support of Deutsche Forschungsgemeinschaft (DFG) through project B06 of SFB 1143: Correlated Magnetism: From Frustration To Topology (ID 247310070). We thank O. Benton, N. Shannon and H. Yan for their insightful discussions. We also thank the sample environment teams of FRM II and ILL for their technical support during the experiments. 
\end{acknowledgments}

\appendix

\label{appendix}

\section{Data treatment}
Inelastic Neutron scattering data were collected on the direct geometry, time-of-flight spectrometer, IN5 \cite{IN5}, at the Intitute Laue Langevin, France. $\omega$-scans over a large angular range were performed with neutrons of wavelength 6~\AA\ and energy resolution 0.041~meV and data were collected above and below the transition temperature $T_{\rm{N}}=0.55~$K, at $T=40~$mK and 600~mK respectively. In addition, data were collected at $T=25$~K over the same wavevector region, with approximately the same statistics as the low temperature datasets. This temperature is well below the first crystal field excited state, which for the ${\rm Nd_2Hf_2O_7}$ is 276.2~K and simultaneously is far above $T_{\rm N}$ \cite{Anand2015, Anand2017} so magnetic correlations should be absent. The data presented in figures 2, 3 and 4 of the main paper have been background-subtracted and normalized. The purpose of the normalization is to remove the effects of rotation-dependent absorption due to the sample and sample environment. The normalization was performed using the 25~K dataset. It was first integrated around zero energy transfer over $E=0\pm0.06$ meV and the signal due to the Bragg peaks was removed to give a wavevector-dependent normalization file. The data at all three temperatures were then normalized by divided by this file. The 25~K dataset was also used to remove the non-magnetic background e.g.\ due to the sample holder and instrument components, from the low temperature datasets. To achieve this the slices through the normalized datasets at 40~mK and 600~mK have had the corresponding slices through the normalized 25~K dataset subtracted from them. In the case of the energy-wavevector slices along the different wavevector directions in Fig.\ 3, the data were symmetrized about various high symmetry planes to improve the statistics.

Polarized neutron diffraction data were collected on the DNS diffractometer at the Meier-Leibniz Zentrum, M\"{u}nich, Germany at 88~mK, 600~mK and 25~K.  The neutrons were polarized along the (vertical) $Z$-direction and both spin-flip ($Z$SF) and non-spin-flip ($Z$NSF) channels were measured. All the data were corrected for flipping ratio and the 25~K dataset was subtracted from the low temperature dataset as a background. 

Further details of the data treatment can be found in Ref.\ \cite{ThesisSamartzis}.

\section{Quantum XYZ model}

The ground state wavefunction of the Nd$^{3+}$ ion is a dipolar-octupolar doublet \cite{Huang2014}. For symmetry reasons, the nearest-neighbor Hamiltonian for the effective pseudospin-1/2 model takes the form \cite{ Huang2014} 
\begin{equation}\label{eq1}
H_{ex}^{DO}=\sum_{<ij>} J_x\tau_i^x\tau_j^x+J_y\tau_i^y\tau_j^y+J_z\tau_i^z\tau_j^z+J_{xz}(\tau_i^x\tau_j^z+\tau_i^z\tau_j^x)
\end{equation}
where the summation is over the nearest neighbors. $\tau_i^{\alpha}$ is the $\alpha$-component of the $i^{th}$ pseudospin-1/2 operator and $J_{\alpha}$ is the exchange interaction between the $\alpha$-components of neighboring spins.
The local coordinate system is used in Eq.~\eqref{eq1} where the local $z$-axis points towards the center of each tetrahedron. The Hamiltonian can be simplified by a rotation about the $y$-axis to a rotated local coordinate system \cite{Benton2016}. The $y$-axis was chosen as the rotation axis due to the D3d symmetry. Under this transformation, the $z$-axis is rotated by $\theta$ into the $\widetilde{z}$-axis and the $\widetilde{x}$-axis is the cross product of the $\widetilde{y}=y$ and $\widetilde{z}$-axes \cite{Ross2011, HanYan2018}.
\begin{equation}\label{trans}
\begin{gathered}
\widetilde{\tau}_i^{\widetilde{x}}=cos(\theta)\tau_i^x+sin(\theta)\tau_i^z,  \\
\widetilde{\tau}_i^{\widetilde{y}}=\tau_i^y\\
\widetilde{\tau}_i^{\widetilde{z}}=cos(\theta)\tau_i^z-sin(\theta)\tau_i^x,  \\
tan(2\theta)=\frac{2J_{xy}}{J_x-J_z}
\end{gathered} 
\end{equation}
Applying this transformation gives us the XYZ Hamiltonian used in the main part of the paper 
\begin{equation}
\label{eq:Hamilt2}
H_{XYZ}^{DO}=\sum_{<ij>} \widetilde{J}_{\widetilde{x}}\widetilde{\tau}_i^{\widetilde{x}}\widetilde{\tau}_j^{\widetilde{x}}+\widetilde{J}_{\widetilde{y}}\widetilde{\tau}_i^{\widetilde{y}}\widetilde{\tau}_j^{\widetilde{y}}+
\widetilde{J}_{\widetilde{z}}\widetilde{\tau}_i^{\widetilde{z}}\widetilde{\tau}_j^{\widetilde{z}}
\end{equation} 
where the transformed exchange constants $\widetilde{J}_{{\widetilde{\alpha}}}$ are 
\begin{equation}\label{trans2}
\begin{gathered}
J_x=\widetilde{J}_{\widetilde{x}}cos^2(\theta)+\widetilde{J}_{\widetilde{z}}sin^2(\theta), \\
J_z=\widetilde{J}_{\widetilde{z}}cos^2(\theta)+\widetilde{J}_{\widetilde{x}}sin^2(\theta) 
\end{gathered} 
\end{equation}
Note that this Hamiltonian is independent of the angle $\theta$. $\theta$ does however play a major role in the susceptibility and the overall neutron scattering intensities, this is because only components of magnetization along the $z$-axis are observable due to the $g$-factor being non-zero in this direction only: $g_{zz}=5.1$, $g_{xx}=g_{yy}=0$.

\section{The spin-wave simulations}

The XYZ Hamiltonian Eq.~\eqref{eq:Hamilt2}, was simulated using linear spin-wave theory within the matlab package, spinW \cite{TothPRX}. SpinW calculations are performed in the global coordinate system, therefore the fitted parameters had to be transformed from the rotated local $\widetilde{x}$, $\widetilde{y}$, $\widetilde{z}$-axes to the local $x$, $y$, $z$-axes and finally to the global coordinate system. To switch between the different coordinate systems, an appropriate transformation matrix was applied \cite{ThesisXu}. 

The simulations shown in figures 2, 3 and 4 of the main paper were convoluted with a gaussian peak of width 0.041 meV to mimic the effects of the instrumental energy resolution for the selected measuring parameters. The magnetic form factor of the Nd$^{3+}$ ion was also included.

\section{Determination of the exchange interactions}

The exchange interactions $\widetilde{J}_{\widetilde{\alpha}}$, of ${\rm Nd_2Hf_2O_7}$ were determined by fitting linear spin-wave theory to the data. The XYZ Hamiltonian (Eq.~\eqref{eq:Hamilt2}) identifies three characteristic energies in the magnetic excitation scheme of $\rm Nd_2Hf_2O_7$. These are \cite{Benton2016}:
\begin{equation} 
\label{eq:gap}
\begin{gathered}
E_0 = \sqrt{(3| \widetilde{J}_{\widetilde{z}}|-\widetilde{J}_{\widetilde{x}})(3| \widetilde{J}_{\widetilde{z}}|-\widetilde{J}_{\widetilde{y}})},  \\
E_1 = \sqrt{(3| \widetilde{J}_{\widetilde{z}}|+\widetilde{J}_{\widetilde{x}})(3|  \widetilde{J}_{\widetilde{z}}|+\widetilde{J}_{\widetilde{y}})}, \\
E_2 = 3\sqrt{(| \widetilde{J}_{\widetilde{z}}|+\widetilde{J}_{\widetilde{x}})(|  \widetilde{J}_{\widetilde{z}}|+\widetilde{J}_{\widetilde{y}})} 
\end{gathered} 
\end{equation} 
where $E_0$ is the energy of the gap to the pinch point mode found from our data to be $E_0=0.094$~meV, $ E_1$ is the position where the two high energy dispersive modes intersect which is $E_1$=0.21~meV (see Fig. 3(a) and 3(c) of the main paper) and finally $E_2$ is the maximum of the spin waves which is $E_2=0.29$~meV (see Fig. 3(c) and 3(d) of the main paper). 

The values of these parameters extracted from the data were used to determine the starting exchange parameters for the fitting. 
The exchange parameters were scanned about these starting values and the corresponding simulations from the spinW program were compared to the data in terms of both the energies of the excitations and their intensities. The fitted parameters are: $\widetilde{J}_{\widetilde{x}}=0.106(5)$~meV, $\widetilde{J}_{\widetilde{y}}=0.008(5)$~meV and $\widetilde{J}_{\widetilde{z}}=-0.057(4)$~meV. Further details of the determination of the exchange interaction can be found in Ref.\ \cite{ThesisSamartzis}.\\

\section{Estimation of the $\theta$ angle}

The angle $\theta$ which is the angle of rotation between the two local coordinate frames, $x$, $y$, $z$ and $\widetilde{x}$, $\widetilde{y}$, $\widetilde{z}$, can be calculated from the Curie-Weiss temperature of ${\rm Nd_2Hf_2O_7}$. According to Ref.\ \cite{Benton2016}, $T_{\rm CW}$ can be expressed in terms of the exchange interactions via 
\begin{equation}
\label{CurieWeiss}
T_{\rm CW}=\frac{1}{2k_{\rm B}}(\widetilde{J}_{\widetilde{z}} \cos^2(\theta)+\widetilde{J}_{\widetilde{x}} \sin^2(\theta)).
\end{equation} 
For ${\rm Nd_2Hf_2O_7}$, $T_{\rm CW}=0.24(2)$~K \cite{Anand2015} which yields $\theta = 72.5^{\circ}$.  This value is significantly larger than the value $\theta = 47.6^{\circ}$ found for ${\rm Nd_2Zr_2O_7}$ \cite{ThesisXu}.

\section{Parameter predictions}

The ordered moment can be calculated from the angle $\theta$ using the equation \cite{Benton2016}.  
\begin{equation} \label{moment}
\begin{gathered}
\begin{aligned}
m_z &= \mu_{\rm B}g_{zz}\langle\tau_{z}\rangle \\
&= \mu_{\rm B}g_{zz}(\langle\widetilde{\tau}_{\widetilde{z}}\rangle \cos(\theta)+\langle\widetilde{\tau}_{\widetilde{x}}\rangle \sin(\theta)) \\
\end{aligned}
\end{gathered}
\end{equation} 
The spins order along the $\widetilde{\tau}_{\widetilde{z}}$-axis, however due to zero point fluctuations we would expect a reduction of $\approx$10\% in the ordered spin, giving $\langle\widetilde{\tau}_{\widetilde{z}}\rangle\approx 0.45$ and $\langle\widetilde{\tau}_{\widetilde{x}}\rangle=0$. Therefore
\begin{equation} \label{moment}
\begin{gathered}
\begin{aligned}
m_z &\approx 5.1*0.45*\cos{0.89}~\mu_{\rm B} \\
&\approx 0.63~\mu_{\rm B}
\end{aligned}
\end{gathered}
\end{equation} 
This value is very similar to the value obtained for  ${\rm Nd_2Hf_2O_7}$ from powder neutron diffraction at $T=0.1$~K of $m \approx 0.62 \mu_{\rm B}$/Nd$^{3+}$, \cite{Anand2015} comfirming the value of $\theta$.

\section{Inelastic neutron scattering data above $T_{\rm N}$}
\setcounter{figure}{0} \renewcommand{\thefigure}{G.\arabic{figure}}
\begin{figure}
	\centering
	\begin{tabular}{c @{\qquad} c }	
		\includegraphics[width=0.95\columnwidth,keepaspectratio]{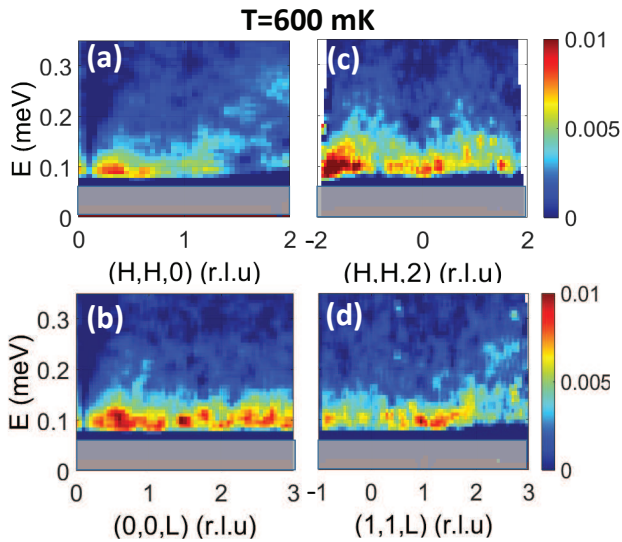} 
	\end{tabular}
	\caption{Energy-wavevector maps of the INS spectrum above $T_{\rm N}$ at 600~mK. The data were integrated over the out-of-plane wavevector and the in-plane wavevector perpendicular to the slice by ${\rm \pm 0.2}$ r.l.u. and they were normalized and background-subtracted. The data below 0.06~meV (shaded region) are unreliable due to background oversubtraction.}
	\label{fig:FigG1}
\end{figure}

Fig.~\ref{fig:FigG1} presents energy-wavevector slices along four high symmetry directions for data collected at 600~mK. A comparison with the corresponding results in Fig.\ 3 of the main paper shows that the signal decreases by almost an order of magnitude compared with that at $T=40$~mK. The signal is broad and diffuse although interestingly it seems to follow the shape of the dispersive excitations in the ordered state. Finally, there is no evidence of the flat mode along any of the $Q$-vectors suggesting that it has moved to lower energies, however the background over-subtraction for $E<0.06$~meV prevents the closure of the gap from being detected. 

\bibliographystyle{apsrev4-1}

%\bibliography{cite1}

\begin{thebibliography}{43}%
	\makeatletter
	\providecommand \@ifxundefined [1]{%
		\@ifx{#1\undefined}
	}%
	\providecommand \@ifnum [1]{%
		\ifnum #1\expandafter \@firstoftwo
		\else \expandafter \@secondoftwo
		\fi
	}%
	\providecommand \@ifx [1]{%
		\ifx #1\expandafter \@firstoftwo
		\else \expandafter \@secondoftwo
		\fi
	}%
	\providecommand \natexlab [1]{#1}%
	\providecommand \enquote  [1]{``#1''}%
	\providecommand \bibnamefont  [1]{#1}%
	\providecommand \bibfnamefont [1]{#1}%
	\providecommand \citenamefont [1]{#1}%
	\providecommand \href@noop [0]{\@secondoftwo}%
	\providecommand \href [0]{\begingroup \@sanitize@url \@href}%
	\providecommand \@href[1]{\@@startlink{#1}\@@href}%
	\providecommand \@@href[1]{\endgroup#1\@@endlink}%
	\providecommand \@sanitize@url [0]{\catcode `\\12\catcode `\$12\catcode
		`\&12\catcode `\#12\catcode `\^12\catcode `\_12\catcode `\%12\relax}%
	\providecommand \@@startlink[1]{}%
	\providecommand \@@endlink[0]{}%
	\providecommand \url  [0]{\begingroup\@sanitize@url \@url }%
	\providecommand \@url [1]{\endgroup\@href {#1}{\urlprefix }}%
	\providecommand \urlprefix  [0]{URL }%
	\providecommand \Eprint [0]{\href }%
	\providecommand \doibase [0]{http://dx.doi.org/}%
	\providecommand \selectlanguage [0]{\@gobble}%
	\providecommand \bibinfo  [0]{\@secondoftwo}%
	\providecommand \bibfield  [0]{\@secondoftwo}%
	\providecommand \translation [1]{[#1]}%
	\providecommand \BibitemOpen [0]{}%
	\providecommand \bibitemStop [0]{}%
	\providecommand \bibitemNoStop [0]{.\EOS\space}%
	\providecommand \EOS [0]{\spacefactor3000\relax}%
	\providecommand \BibitemShut  [1]{\csname bibitem#1\endcsname}%
	\let\auto@bib@innerbib\@empty
	%</preamble>
	\bibitem [{\citenamefont {Henley}(2010)}]{Henley}%
	\BibitemOpen
	\bibfield  {author} {\bibinfo {author} {\bibfnamefont {C.~L.}\ \bibnamefont
			{Henley}},\ }\href {\doibase 10.1146/annurev-conmatphys-070909-104138}
	{\bibfield  {journal} {\bibinfo  {journal} {Annual Review of Condensed Matter
				Physics}\ }\textbf {\bibinfo {volume} {1}},\ \bibinfo {pages} {179} (\bibinfo
		{year} {2010})},\ \Eprint
	{http://arxiv.org/abs/https://doi.org/10.1146/annurev-conmatphys-070909-104138}
	{https://doi.org/10.1146/annurev-conmatphys-070909-104138} \BibitemShut
	{NoStop}%
	\bibitem [{\citenamefont {Bramwell}\ and\ \citenamefont
		{Gingras}(2001)}]{Bramwell2001}%
	\BibitemOpen
	\bibfield  {author} {\bibinfo {author} {\bibfnamefont {S.~T.}\ \bibnamefont
			{Bramwell}}\ and\ \bibinfo {author} {\bibfnamefont {M.~J.~P.}\ \bibnamefont
			{Gingras}},\ }\href {\doibase 10.1126/science.1064761} {\bibfield  {journal}
		{\bibinfo  {journal} {Science}\ }\textbf {\bibinfo {volume} {294}},\ \bibinfo
		{pages} {1495} (\bibinfo {year} {2001})}\BibitemShut {NoStop}%
	\bibitem [{\citenamefont {Fennell}\ \emph {et~al.}(2009)\citenamefont
		{Fennell}, \citenamefont {Deen}, \citenamefont {Wildes}, \citenamefont
		{Schmalzl}, \citenamefont {Prabhakaran}, \citenamefont {Boothroyd},
		\citenamefont {Aldus}, \citenamefont {McMorrow},\ and\ \citenamefont
		{Bramwell}}]{Fennell2009}%
	\BibitemOpen
	\bibfield  {author} {\bibinfo {author} {\bibfnamefont {T.}~\bibnamefont
			{Fennell}}, \bibinfo {author} {\bibfnamefont {P.~P.}\ \bibnamefont {Deen}},
		\bibinfo {author} {\bibfnamefont {A.~R.}\ \bibnamefont {Wildes}}, \bibinfo
		{author} {\bibfnamefont {K.}~\bibnamefont {Schmalzl}}, \bibinfo {author}
		{\bibfnamefont {D.}~\bibnamefont {Prabhakaran}}, \bibinfo {author}
		{\bibfnamefont {A.~T.}\ \bibnamefont {Boothroyd}}, \bibinfo {author}
		{\bibfnamefont {R.~J.}\ \bibnamefont {Aldus}}, \bibinfo {author}
		{\bibfnamefont {D.~F.}\ \bibnamefont {McMorrow}}, \ and\ \bibinfo {author}
		{\bibfnamefont {S.~T.}\ \bibnamefont {Bramwell}},\ }\href {\doibase
		10.1126/science.1177582} {\bibfield  {journal} {\bibinfo  {journal}
			{Science}\ }\textbf {\bibinfo {volume} {326}},\ \bibinfo {pages} {415}
		(\bibinfo {year} {2009})}\BibitemShut {NoStop}%
	\bibitem [{\citenamefont {Morris}\ \emph {et~al.}(2009)\citenamefont {Morris},
		\citenamefont {Tennant}, \citenamefont {Grigera}, \citenamefont {Klemke},
		\citenamefont {Castelnovo}, \citenamefont {Moessner}, \citenamefont
		{Czternasty}, \citenamefont {Meissner}, \citenamefont {Rule}, \citenamefont
		{Hoffmann}, \citenamefont {Kiefer}, \citenamefont {Gerischer}, \citenamefont
		{Slobinsky},\ and\ \citenamefont {Perry}}]{morris2009}%
	\BibitemOpen
	\bibfield  {author} {\bibinfo {author} {\bibfnamefont {D.~J.~P.}\
			\bibnamefont {Morris}}, \bibinfo {author} {\bibfnamefont {D.~A.}\
			\bibnamefont {Tennant}}, \bibinfo {author} {\bibfnamefont {S.~A.}\
			\bibnamefont {Grigera}}, \bibinfo {author} {\bibfnamefont {B.}~\bibnamefont
			{Klemke}}, \bibinfo {author} {\bibfnamefont {C.}~\bibnamefont {Castelnovo}},
		\bibinfo {author} {\bibfnamefont {R.}~\bibnamefont {Moessner}}, \bibinfo
		{author} {\bibfnamefont {C.}~\bibnamefont {Czternasty}}, \bibinfo {author}
		{\bibfnamefont {M.}~\bibnamefont {Meissner}}, \bibinfo {author}
		{\bibfnamefont {K.~C.}\ \bibnamefont {Rule}}, \bibinfo {author}
		{\bibfnamefont {J.-U.}\ \bibnamefont {Hoffmann}}, \bibinfo {author}
		{\bibfnamefont {K.}~\bibnamefont {Kiefer}}, \bibinfo {author} {\bibfnamefont
			{S.}~\bibnamefont {Gerischer}}, \bibinfo {author} {\bibfnamefont
			{D.}~\bibnamefont {Slobinsky}}, \ and\ \bibinfo {author} {\bibfnamefont
			{R.~S.}\ \bibnamefont {Perry}},\ }\href {\doibase 10.1126/science.1178868}
	{\bibfield  {journal} {\bibinfo  {journal} {Science}\ }\textbf {\bibinfo
			{volume} {326}},\ \bibinfo {pages} {411} (\bibinfo {year}
		{2009})}\BibitemShut {NoStop}%
	\bibitem [{\citenamefont {den Hertog}\ and\ \citenamefont
		{Gingras}(2000)}]{Hertog2000}%
	\BibitemOpen
	\bibfield  {author} {\bibinfo {author} {\bibfnamefont {B.~C.}\ \bibnamefont
			{den Hertog}}\ and\ \bibinfo {author} {\bibfnamefont {M.~J.~P.}\ \bibnamefont
			{Gingras}},\ }\href {\doibase 10.1103/PhysRevLett.84.3430} {\bibfield
		{journal} {\bibinfo  {journal} {Phys. Rev. Lett.}\ }\textbf {\bibinfo
			{volume} {84}},\ \bibinfo {pages} {3430} (\bibinfo {year}
		{2000})}\BibitemShut {NoStop}%
	\bibitem [{\citenamefont {Castelnovo}\ \emph {et~al.}(2008)\citenamefont
		{Castelnovo}, \citenamefont {Moessner},\ and\ \citenamefont
		{Sondhi}}]{castelnovo2008}%
	\BibitemOpen
	\bibfield  {author} {\bibinfo {author} {\bibfnamefont {C.}~\bibnamefont
			{Castelnovo}}, \bibinfo {author} {\bibfnamefont {R.}~\bibnamefont
			{Moessner}}, \ and\ \bibinfo {author} {\bibfnamefont {S.~L.}\ \bibnamefont
			{Sondhi}},\ }\href {\doibase 10.1038/nature06433} {\bibfield  {journal}
		{\bibinfo  {journal} {Nature}\ }\textbf {\bibinfo {volume} {451}},\ \bibinfo
		{pages} {42} (\bibinfo {year} {2008})}\BibitemShut {NoStop}%
	\bibitem [{\citenamefont {Hermele}\ \emph {et~al.}(2004)\citenamefont
		{Hermele}, \citenamefont {Fisher},\ and\ \citenamefont
		{Balents}}]{Hermele2004}%
	\BibitemOpen
	\bibfield  {author} {\bibinfo {author} {\bibfnamefont {M.}~\bibnamefont
			{Hermele}}, \bibinfo {author} {\bibfnamefont {M.~P.~A.}\ \bibnamefont
			{Fisher}}, \ and\ \bibinfo {author} {\bibfnamefont {L.}~\bibnamefont
			{Balents}},\ }\href {\doibase 10.1103/PhysRevB.69.064404} {\bibfield
		{journal} {\bibinfo  {journal} {Phys. Rev. B}\ }\textbf {\bibinfo {volume}
			{69}},\ \bibinfo {pages} {064404} (\bibinfo {year} {2004})}\BibitemShut
	{NoStop}%
	\bibitem [{\citenamefont {Shannon}\ \emph {et~al.}(2012)\citenamefont
		{Shannon}, \citenamefont {Sikora}, \citenamefont {Pollmann}, \citenamefont
		{Penc},\ and\ \citenamefont {Fulde}}]{Shannon2012}%
	\BibitemOpen
	\bibfield  {author} {\bibinfo {author} {\bibfnamefont {N.}~\bibnamefont
			{Shannon}}, \bibinfo {author} {\bibfnamefont {O.}~\bibnamefont {Sikora}},
		\bibinfo {author} {\bibfnamefont {F.}~\bibnamefont {Pollmann}}, \bibinfo
		{author} {\bibfnamefont {K.}~\bibnamefont {Penc}}, \ and\ \bibinfo {author}
		{\bibfnamefont {P.}~\bibnamefont {Fulde}},\ }\href {\doibase
		10.1103/PhysRevLett.108.067204} {\bibfield  {journal} {\bibinfo  {journal}
			{Phys. Rev. Lett.}\ }\textbf {\bibinfo {volume} {108}},\ \bibinfo {pages}
		{067204} (\bibinfo {year} {2012})}\BibitemShut {NoStop}%
	\bibitem [{\citenamefont {Benton}\ \emph {et~al.}(2012)\citenamefont {Benton},
		\citenamefont {Sikora},\ and\ \citenamefont {Shannon}}]{Benton2012}%
	\BibitemOpen
	\bibfield  {author} {\bibinfo {author} {\bibfnamefont {O.}~\bibnamefont
			{Benton}}, \bibinfo {author} {\bibfnamefont {O.}~\bibnamefont {Sikora}}, \
		and\ \bibinfo {author} {\bibfnamefont {N.}~\bibnamefont {Shannon}},\ }\href
	{\doibase 10.1103/PhysRevB.86.075154} {\bibfield  {journal} {\bibinfo
			{journal} {Phys. Rev. B}\ }\textbf {\bibinfo {volume} {86}},\ \bibinfo
		{pages} {075154} (\bibinfo {year} {2012})}\BibitemShut {NoStop}%
	\bibitem [{\citenamefont {Yan}\ \emph {et~al.}(2018)\citenamefont {Yan},
		\citenamefont {Pohle},\ and\ \citenamefont {Shannon}}]{HanYan2018}%
	\BibitemOpen
	\bibfield  {author} {\bibinfo {author} {\bibfnamefont {H.}~\bibnamefont
			{Yan}}, \bibinfo {author} {\bibfnamefont {R.}~\bibnamefont {Pohle}}, \ and\
		\bibinfo {author} {\bibfnamefont {N.}~\bibnamefont {Shannon}},\ }\href
	{\doibase 10.1103/PhysRevB.98.140402} {\bibfield  {journal} {\bibinfo
			{journal} {Phys. Rev. B}\ }\textbf {\bibinfo {volume} {98}},\ \bibinfo
		{pages} {140402(R)} (\bibinfo {year} {2018})}\BibitemShut {NoStop}%
	\bibitem [{\citenamefont {Benton}(2016)}]{Benton2016}%
	\BibitemOpen
	\bibfield  {author} {\bibinfo {author} {\bibfnamefont {O.}~\bibnamefont
			{Benton}},\ }\href {\doibase 10.1103/PhysRevB.94.104430} {\bibfield
		{journal} {\bibinfo  {journal} {Phys. Rev. B}\ }\textbf {\bibinfo {volume}
			{94}},\ \bibinfo {pages} {104430} (\bibinfo {year} {2016})}\BibitemShut
	{NoStop}%
	\bibitem [{SM()}]{SM}%
	\BibitemOpen
	\href@noop {} {\ }\BibitemShut {NoStop}%
	\bibitem [{\citenamefont {Brooks-Bartlett}\ \emph {et~al.}(2014)\citenamefont
		{Brooks-Bartlett}, \citenamefont {Banks}, \citenamefont {Jaubert},
		\citenamefont {Harman-Clarke},\ and\ \citenamefont {Holdsworth}}]{MomFrag}%
	\BibitemOpen
	\bibfield  {author} {\bibinfo {author} {\bibfnamefont {M.~E.}\ \bibnamefont
			{Brooks-Bartlett}}, \bibinfo {author} {\bibfnamefont {S.~T.}\ \bibnamefont
			{Banks}}, \bibinfo {author} {\bibfnamefont {L.~D.~C.}\ \bibnamefont
			{Jaubert}}, \bibinfo {author} {\bibfnamefont {A.}~\bibnamefont
			{Harman-Clarke}}, \ and\ \bibinfo {author} {\bibfnamefont {P.~C.~W.}\
			\bibnamefont {Holdsworth}},\ }\href {\doibase 10.1103/PhysRevX.4.011007}
	{\bibfield  {journal} {\bibinfo  {journal} {Phys. Rev. X}\ }\textbf {\bibinfo
			{volume} {4}},\ \bibinfo {pages} {011007} (\bibinfo {year}
		{2014})}\BibitemShut {NoStop}%
	\bibitem [{\citenamefont {Xu}\ \emph {et~al.}(2015)\citenamefont {Xu},
		\citenamefont {Anand}, \citenamefont {Bera}, \citenamefont {Frontzek},
		\citenamefont {Abernathy}, \citenamefont {Casati}, \citenamefont
		{Siemensmeyer},\ and\ \citenamefont {Lake}}]{Xu2015}%
	\BibitemOpen
	\bibfield  {author} {\bibinfo {author} {\bibfnamefont {J.}~\bibnamefont
			{Xu}}, \bibinfo {author} {\bibfnamefont {V.~K.}\ \bibnamefont {Anand}},
		\bibinfo {author} {\bibfnamefont {A.~K.}\ \bibnamefont {Bera}}, \bibinfo
		{author} {\bibfnamefont {M.}~\bibnamefont {Frontzek}}, \bibinfo {author}
		{\bibfnamefont {D.~L.}\ \bibnamefont {Abernathy}}, \bibinfo {author}
		{\bibfnamefont {N.}~\bibnamefont {Casati}}, \bibinfo {author} {\bibfnamefont
			{K.}~\bibnamefont {Siemensmeyer}}, \ and\ \bibinfo {author} {\bibfnamefont
			{B.}~\bibnamefont {Lake}},\ }\href {\doibase 10.1103/PhysRevB.92.224430}
	{\bibfield  {journal} {\bibinfo  {journal} {Phys. Rev. B}\ }\textbf {\bibinfo
			{volume} {92}},\ \bibinfo {pages} {224430} (\bibinfo {year}
		{2015})}\BibitemShut {NoStop}%
	\bibitem [{\citenamefont {Lhotel}\ \emph {et~al.}(2015)\citenamefont {Lhotel},
		\citenamefont {Petit}, \citenamefont {Guitteny}, \citenamefont {Florea},
		\citenamefont {Ciomaga~Hatnean}, \citenamefont {Colin}, \citenamefont
		{Ressouche}, \citenamefont {Lees},\ and\ \citenamefont
		{Balakrishnan}}]{Lhotel2015}%
	\BibitemOpen
	\bibfield  {author} {\bibinfo {author} {\bibfnamefont {E.}~\bibnamefont
			{Lhotel}}, \bibinfo {author} {\bibfnamefont {S.}~\bibnamefont {Petit}},
		\bibinfo {author} {\bibfnamefont {S.}~\bibnamefont {Guitteny}}, \bibinfo
		{author} {\bibfnamefont {O.}~\bibnamefont {Florea}}, \bibinfo {author}
		{\bibfnamefont {M.}~\bibnamefont {Ciomaga~Hatnean}}, \bibinfo {author}
		{\bibfnamefont {C.}~\bibnamefont {Colin}}, \bibinfo {author} {\bibfnamefont
			{E.}~\bibnamefont {Ressouche}}, \bibinfo {author} {\bibfnamefont {M.~R.}\
			\bibnamefont {Lees}}, \ and\ \bibinfo {author} {\bibfnamefont
			{G.}~\bibnamefont {Balakrishnan}},\ }\href {\doibase
		10.1103/PhysRevLett.115.197202} {\bibfield  {journal} {\bibinfo  {journal}
			{Phys. Rev. Lett.}\ }\textbf {\bibinfo {volume} {115}},\ \bibinfo {pages}
		{197202} (\bibinfo {year} {2015})}\BibitemShut {NoStop}%
	\bibitem [{\citenamefont {Xu}\ \emph {et~al.}(2016)\citenamefont {Xu},
		\citenamefont {Balz}, \citenamefont {Baines}, \citenamefont {Luetkens},\ and\
		\citenamefont {Lake}}]{Xu2016}%
	\BibitemOpen
	\bibfield  {author} {\bibinfo {author} {\bibfnamefont {J.}~\bibnamefont
			{Xu}}, \bibinfo {author} {\bibfnamefont {C.}~\bibnamefont {Balz}}, \bibinfo
		{author} {\bibfnamefont {C.}~\bibnamefont {Baines}}, \bibinfo {author}
		{\bibfnamefont {H.}~\bibnamefont {Luetkens}}, \ and\ \bibinfo {author}
		{\bibfnamefont {B.}~\bibnamefont {Lake}},\ }\href {\doibase
		10.1103/PhysRevB.94.064425} {\bibfield  {journal} {\bibinfo  {journal} {Phys.
				Rev. B}\ }\textbf {\bibinfo {volume} {94}},\ \bibinfo {pages} {064425}
		(\bibinfo {year} {2016})}\BibitemShut {NoStop}%
	\bibitem [{\citenamefont {Petit}\ \emph {et~al.}(2016)\citenamefont {Petit},
		\citenamefont {Lhotel}, \citenamefont {Canals}, \citenamefont
		{Ciomaga~Hatnean}, \citenamefont {Ollivier}, \citenamefont {Mutka},
		\citenamefont {Ressouche}, \citenamefont {Wildes}, \citenamefont {Lees},\
		and\ \citenamefont {Balakrishnan}}]{Petit2016}%
	\BibitemOpen
	\bibfield  {author} {\bibinfo {author} {\bibfnamefont {S.}~\bibnamefont
			{Petit}}, \bibinfo {author} {\bibfnamefont {E.}~\bibnamefont {Lhotel}},
		\bibinfo {author} {\bibfnamefont {B.}~\bibnamefont {Canals}}, \bibinfo
		{author} {\bibfnamefont {M.}~\bibnamefont {Ciomaga~Hatnean}}, \bibinfo
		{author} {\bibfnamefont {J.}~\bibnamefont {Ollivier}}, \bibinfo {author}
		{\bibfnamefont {H.}~\bibnamefont {Mutka}}, \bibinfo {author} {\bibfnamefont
			{E.}~\bibnamefont {Ressouche}}, \bibinfo {author} {\bibfnamefont {A.~R.}\
			\bibnamefont {Wildes}}, \bibinfo {author} {\bibfnamefont {M.~R.}\
			\bibnamefont {Lees}}, \ and\ \bibinfo {author} {\bibfnamefont
			{G.}~\bibnamefont {Balakrishnan}},\ }\href {\doibase 10.1038/nphys3710}
	{\bibfield  {journal} {\bibinfo  {journal} {Nature Physics}\ }\textbf
		{\bibinfo {volume} {12}},\ \bibinfo {pages} {746} (\bibinfo {year}
		{2016})}\BibitemShut {NoStop}%
	\bibitem [{\citenamefont {Xu}\ \emph {et~al.}(2019)\citenamefont {Xu},
		\citenamefont {Benton}, \citenamefont {Anand}, \citenamefont {Islam},
		\citenamefont {Guidi}, \citenamefont {Ehlers}, \citenamefont {Feng},
		\citenamefont {Su}, \citenamefont {Sakai}, \citenamefont {Gegenwart},\ and\
		\citenamefont {Lake}}]{Xu2019}%
	\BibitemOpen
	\bibfield  {author} {\bibinfo {author} {\bibfnamefont {J.}~\bibnamefont
			{Xu}}, \bibinfo {author} {\bibfnamefont {O.}~\bibnamefont {Benton}}, \bibinfo
		{author} {\bibfnamefont {V.~K.}\ \bibnamefont {Anand}}, \bibinfo {author}
		{\bibfnamefont {A.~T. M.~N.}\ \bibnamefont {Islam}}, \bibinfo {author}
		{\bibfnamefont {T.}~\bibnamefont {Guidi}}, \bibinfo {author} {\bibfnamefont
			{G.}~\bibnamefont {Ehlers}}, \bibinfo {author} {\bibfnamefont
			{E.}~\bibnamefont {Feng}}, \bibinfo {author} {\bibfnamefont {Y.}~\bibnamefont
			{Su}}, \bibinfo {author} {\bibfnamefont {A.}~\bibnamefont {Sakai}}, \bibinfo
		{author} {\bibfnamefont {P.}~\bibnamefont {Gegenwart}}, \ and\ \bibinfo
		{author} {\bibfnamefont {B.}~\bibnamefont {Lake}},\ }\href {\doibase
		10.1103/PhysRevB.99.144420} {\bibfield  {journal} {\bibinfo  {journal} {Phys.
				Rev. B}\ }\textbf {\bibinfo {volume} {99}},\ \bibinfo {pages} {144420}
		(\bibinfo {year} {2019})}\BibitemShut {NoStop}%
	\bibitem [{\citenamefont {Lhotel}\ \emph {et~al.}(2018)\citenamefont {Lhotel},
		\citenamefont {Petit}, \citenamefont {Ciomaga~Hatnean}, \citenamefont
		{Ollivier}, \citenamefont {Mutka}, \citenamefont {Ressouche}, \citenamefont
		{Lees},\ and\ \citenamefont {Balakrishnan}}]{Lhotel2018}%
	\BibitemOpen
	\bibfield  {author} {\bibinfo {author} {\bibfnamefont {E.}~\bibnamefont
			{Lhotel}}, \bibinfo {author} {\bibfnamefont {S.}~\bibnamefont {Petit}},
		\bibinfo {author} {\bibfnamefont {M.}~\bibnamefont {Ciomaga~Hatnean}},
		\bibinfo {author} {\bibfnamefont {J.}~\bibnamefont {Ollivier}}, \bibinfo
		{author} {\bibfnamefont {H.}~\bibnamefont {Mutka}}, \bibinfo {author}
		{\bibfnamefont {E.}~\bibnamefont {Ressouche}}, \bibinfo {author}
		{\bibfnamefont {M.~R.}\ \bibnamefont {Lees}}, \ and\ \bibinfo {author}
		{\bibfnamefont {G.}~\bibnamefont {Balakrishnan}},\ }\href {\doibase
		10.1038/s41467-018-06212-2} {\bibfield  {journal} {\bibinfo  {journal}
			{Nature Communications}\ }\textbf {\bibinfo {volume} {9}},\ \bibinfo {pages}
		{3786} (\bibinfo {year} {2018})}\BibitemShut {NoStop}%
	\bibitem [{\citenamefont {Xu}\ \emph {et~al.}(2018)\citenamefont {Xu},
		\citenamefont {Islam}, \citenamefont {Glavatskyy}, \citenamefont {Reehuis},
		\citenamefont {Hoffmann},\ and\ \citenamefont {Lake}}]{Xu2018}%
	\BibitemOpen
	\bibfield  {author} {\bibinfo {author} {\bibfnamefont {J.}~\bibnamefont
			{Xu}}, \bibinfo {author} {\bibfnamefont {A.~T. M.~N.}\ \bibnamefont {Islam}},
		\bibinfo {author} {\bibfnamefont {I.~N.}\ \bibnamefont {Glavatskyy}},
		\bibinfo {author} {\bibfnamefont {M.}~\bibnamefont {Reehuis}}, \bibinfo
		{author} {\bibfnamefont {J.-U.}\ \bibnamefont {Hoffmann}}, \ and\ \bibinfo
		{author} {\bibfnamefont {B.}~\bibnamefont {Lake}},\ }\href {\doibase
		10.1103/PhysRevB.98.060408} {\bibfield  {journal} {\bibinfo  {journal} {Phys.
				Rev. B}\ }\textbf {\bibinfo {volume} {98}},\ \bibinfo {pages} {060408(R)}
		(\bibinfo {year} {2018})}\BibitemShut {NoStop}%
	\bibitem [{\citenamefont {Opherden}\ \emph {et~al.}(2017)\citenamefont
		{Opherden}, \citenamefont {Hornung}, \citenamefont {Herrmannsd\"orfer},
		\citenamefont {Xu}, \citenamefont {Islam}, \citenamefont {Lake},\ and\
		\citenamefont {Wosnitza}}]{Opherden2017}%
	\BibitemOpen
	\bibfield  {author} {\bibinfo {author} {\bibfnamefont {L.}~\bibnamefont
			{Opherden}}, \bibinfo {author} {\bibfnamefont {J.}~\bibnamefont {Hornung}},
		\bibinfo {author} {\bibfnamefont {T.}~\bibnamefont {Herrmannsd\"orfer}},
		\bibinfo {author} {\bibfnamefont {J.}~\bibnamefont {Xu}}, \bibinfo {author}
		{\bibfnamefont {A.~T. M.~N.}\ \bibnamefont {Islam}}, \bibinfo {author}
		{\bibfnamefont {B.}~\bibnamefont {Lake}}, \ and\ \bibinfo {author}
		{\bibfnamefont {J.}~\bibnamefont {Wosnitza}},\ }\href {\doibase
		10.1103/PhysRevB.95.184418} {\bibfield  {journal} {\bibinfo  {journal} {Phys.
				Rev. B}\ }\textbf {\bibinfo {volume} {95}},\ \bibinfo {pages} {184418}
		(\bibinfo {year} {2017})}\BibitemShut {NoStop}%
	\bibitem [{\citenamefont {Guitteny}\ \emph {et~al.}(2013)\citenamefont
		{Guitteny}, \citenamefont {Robert}, \citenamefont {Bonville}, \citenamefont
		{Ollivier}, \citenamefont {Decorse}, \citenamefont {Steffens}, \citenamefont
		{Boehm}, \citenamefont {Mutka}, \citenamefont {Mirebeau},\ and\ \citenamefont
		{Petit}}]{Guitteny2013}%
	\BibitemOpen
	\bibfield  {author} {\bibinfo {author} {\bibfnamefont {S.}~\bibnamefont
			{Guitteny}}, \bibinfo {author} {\bibfnamefont {J.}~\bibnamefont {Robert}},
		\bibinfo {author} {\bibfnamefont {P.}~\bibnamefont {Bonville}}, \bibinfo
		{author} {\bibfnamefont {J.}~\bibnamefont {Ollivier}}, \bibinfo {author}
		{\bibfnamefont {C.}~\bibnamefont {Decorse}}, \bibinfo {author} {\bibfnamefont
			{P.}~\bibnamefont {Steffens}}, \bibinfo {author} {\bibfnamefont
			{M.}~\bibnamefont {Boehm}}, \bibinfo {author} {\bibfnamefont
			{H.}~\bibnamefont {Mutka}}, \bibinfo {author} {\bibfnamefont
			{I.}~\bibnamefont {Mirebeau}}, \ and\ \bibinfo {author} {\bibfnamefont
			{S.}~\bibnamefont {Petit}},\ }\href {\doibase 10.1103/PhysRevLett.111.087201}
	{\bibfield  {journal} {\bibinfo  {journal} {Phys. Rev. Lett.}\ }\textbf
		{\bibinfo {volume} {111}},\ \bibinfo {pages} {087201} (\bibinfo {year}
		{2013})}\BibitemShut {NoStop}%
	\bibitem [{\citenamefont {Fennell}\ \emph {et~al.}(2014)\citenamefont
		{Fennell}, \citenamefont {Kenzelmann}, \citenamefont {Roessli}, \citenamefont
		{Mutka}, \citenamefont {Ollivier}, \citenamefont {Ruminy}, \citenamefont
		{Stuhr}, \citenamefont {Zaharko}, \citenamefont {Bovo}, \citenamefont
		{Cervellino}, \citenamefont {Haas},\ and\ \citenamefont
		{Cava}}]{Fennell2014}%
	\BibitemOpen
	\bibfield  {author} {\bibinfo {author} {\bibfnamefont {T.}~\bibnamefont
			{Fennell}}, \bibinfo {author} {\bibfnamefont {M.}~\bibnamefont {Kenzelmann}},
		\bibinfo {author} {\bibfnamefont {B.}~\bibnamefont {Roessli}}, \bibinfo
		{author} {\bibfnamefont {H.}~\bibnamefont {Mutka}}, \bibinfo {author}
		{\bibfnamefont {J.}~\bibnamefont {Ollivier}}, \bibinfo {author}
		{\bibfnamefont {M.}~\bibnamefont {Ruminy}}, \bibinfo {author} {\bibfnamefont
			{U.}~\bibnamefont {Stuhr}}, \bibinfo {author} {\bibfnamefont
			{O.}~\bibnamefont {Zaharko}}, \bibinfo {author} {\bibfnamefont
			{L.}~\bibnamefont {Bovo}}, \bibinfo {author} {\bibfnamefont {A.}~\bibnamefont
			{Cervellino}}, \bibinfo {author} {\bibfnamefont {M.~K.}\ \bibnamefont
			{Haas}}, \ and\ \bibinfo {author} {\bibfnamefont {R.~J.}\ \bibnamefont
			{Cava}},\ }\href {\doibase 10.1103/PhysRevLett.112.017203} {\bibfield
		{journal} {\bibinfo  {journal} {Phys. Rev. Lett.}\ }\textbf {\bibinfo
			{volume} {112}},\ \bibinfo {pages} {017203} (\bibinfo {year}
		{2014})}\BibitemShut {NoStop}%
	\bibitem [{\citenamefont {Taillefumier}\ \emph {et~al.}(2014)\citenamefont
		{Taillefumier}, \citenamefont {Robert}, \citenamefont {Henley}, \citenamefont
		{Moessner},\ and\ \citenamefont {Canals}}]{Taillefumier2014}%
	\BibitemOpen
	\bibfield  {author} {\bibinfo {author} {\bibfnamefont {M.}~\bibnamefont
			{Taillefumier}}, \bibinfo {author} {\bibfnamefont {J.}~\bibnamefont
			{Robert}}, \bibinfo {author} {\bibfnamefont {C.~L.}\ \bibnamefont {Henley}},
		\bibinfo {author} {\bibfnamefont {R.}~\bibnamefont {Moessner}}, \ and\
		\bibinfo {author} {\bibfnamefont {B.}~\bibnamefont {Canals}},\ }\href
	{\doibase 10.1103/PhysRevB.90.064419} {\bibfield  {journal} {\bibinfo
			{journal} {Phys. Rev. B}\ }\textbf {\bibinfo {volume} {90}},\ \bibinfo
		{pages} {064419} (\bibinfo {year} {2014})}\BibitemShut {NoStop}%
	\bibitem [{\citenamefont {Robert}\ \emph {et~al.}(2008)\citenamefont {Robert},
		\citenamefont {Canals}, \citenamefont {Simonet},\ and\ \citenamefont
		{Ballou}}]{Robert2008}%
	\BibitemOpen
	\bibfield  {author} {\bibinfo {author} {\bibfnamefont {J.}~\bibnamefont
			{Robert}}, \bibinfo {author} {\bibfnamefont {B.}~\bibnamefont {Canals}},
		\bibinfo {author} {\bibfnamefont {V.}~\bibnamefont {Simonet}}, \ and\
		\bibinfo {author} {\bibfnamefont {R.}~\bibnamefont {Ballou}},\ }\href
	{\doibase 10.1103/PhysRevLett.101.117207} {\bibfield  {journal} {\bibinfo
			{journal} {Phys. Rev. Lett.}\ }\textbf {\bibinfo {volume} {101}},\ \bibinfo
		{pages} {117207} (\bibinfo {year} {2008})}\BibitemShut {NoStop}%
	\bibitem [{\citenamefont {Rau}\ and\ \citenamefont {Gingras}(2016)}]{Rau2016}%
	\BibitemOpen
	\bibfield  {author} {\bibinfo {author} {\bibfnamefont {J.~G.}\ \bibnamefont
			{Rau}}\ and\ \bibinfo {author} {\bibfnamefont {M.~J.~P.}\ \bibnamefont
			{Gingras}},\ }\href {\doibase 10.1038/ncomms12234} {\bibfield  {journal}
		{\bibinfo  {journal} {Nature Communications}\ }\textbf {\bibinfo {volume}
			{7}},\ \bibinfo {pages} {12234} (\bibinfo {year} {2016})}\BibitemShut
	{NoStop}%
	\bibitem [{\citenamefont {Udagawa}\ \emph {et~al.}(2016)\citenamefont
		{Udagawa}, \citenamefont {Jaubert}, \citenamefont {Castelnovo},\ and\
		\citenamefont {Moessner}}]{Udagawa2016}%
	\BibitemOpen
	\bibfield  {author} {\bibinfo {author} {\bibfnamefont {M.}~\bibnamefont
			{Udagawa}}, \bibinfo {author} {\bibfnamefont {L.~D.~C.}\ \bibnamefont
			{Jaubert}}, \bibinfo {author} {\bibfnamefont {C.}~\bibnamefont {Castelnovo}},
		\ and\ \bibinfo {author} {\bibfnamefont {R.}~\bibnamefont {Moessner}},\
	}\href {\doibase 10.1103/PhysRevB.94.104416} {\bibfield  {journal} {\bibinfo
			{journal} {Phys. Rev. B}\ }\textbf {\bibinfo {volume} {94}},\ \bibinfo
		{pages} {104416} (\bibinfo {year} {2016})}\BibitemShut {NoStop}%
	\bibitem [{\citenamefont {Mizoguchi}\ \emph {et~al.}(2017)\citenamefont
		{Mizoguchi}, \citenamefont {Jaubert},\ and\ \citenamefont
		{Udagawa}}]{Mizoguchi2017}%
	\BibitemOpen
	\bibfield  {author} {\bibinfo {author} {\bibfnamefont {T.}~\bibnamefont
			{Mizoguchi}}, \bibinfo {author} {\bibfnamefont {L.~D.~C.}\ \bibnamefont
			{Jaubert}}, \ and\ \bibinfo {author} {\bibfnamefont {M.}~\bibnamefont
			{Udagawa}},\ }\href {\doibase 10.1103/PhysRevLett.119.077207} {\bibfield
		{journal} {\bibinfo  {journal} {Phys. Rev. Lett.}\ }\textbf {\bibinfo
			{volume} {119}},\ \bibinfo {pages} {077207} (\bibinfo {year}
		{2017})}\BibitemShut {NoStop}%
	\bibitem [{\citenamefont {Saha}\ \emph {et~al.}(2021)\citenamefont {Saha},
		\citenamefont {Zhang}, \citenamefont {Lee},\ and\ \citenamefont
		{Chern}}]{Saha2021}%
	\BibitemOpen
	\bibfield  {author} {\bibinfo {author} {\bibfnamefont {P.}~\bibnamefont
			{Saha}}, \bibinfo {author} {\bibfnamefont {D.}~\bibnamefont {Zhang}},
		\bibinfo {author} {\bibfnamefont {S.-H.}\ \bibnamefont {Lee}}, \ and\
		\bibinfo {author} {\bibfnamefont {G.-W.}\ \bibnamefont {Chern}},\ }\href
	{\doibase 10.1103/PhysRevB.103.224402} {\bibfield  {journal} {\bibinfo
			{journal} {Phys. Rev. B}\ }\textbf {\bibinfo {volume} {103}},\ \bibinfo
		{pages} {224402} (\bibinfo {year} {2021})}\BibitemShut {NoStop}%
	\bibitem [{\citenamefont {Mizoguchi}\ \emph {et~al.}(2018)\citenamefont
		{Mizoguchi}, \citenamefont {Jaubert}, \citenamefont {Moessner},\ and\
		\citenamefont {Udagawa}}]{Mizoguchi2018}%
	\BibitemOpen
	\bibfield  {author} {\bibinfo {author} {\bibfnamefont {T.}~\bibnamefont
			{Mizoguchi}}, \bibinfo {author} {\bibfnamefont {L.~D.~C.}\ \bibnamefont
			{Jaubert}}, \bibinfo {author} {\bibfnamefont {R.}~\bibnamefont {Moessner}}, \
		and\ \bibinfo {author} {\bibfnamefont {M.}~\bibnamefont {Udagawa}},\ }\href
	{\doibase 10.1103/PhysRevB.98.144446} {\bibfield  {journal} {\bibinfo
			{journal} {Phys. Rev. B}\ }\textbf {\bibinfo {volume} {98}},\ \bibinfo
		{pages} {144446} (\bibinfo {year} {2018})}\BibitemShut {NoStop}%
	\bibitem [{\citenamefont {Anand}\ \emph {et~al.}(2017)\citenamefont {Anand},
		\citenamefont {Abernathy}, \citenamefont {Adroja}, \citenamefont {Hillier},
		\citenamefont {Biswas},\ and\ \citenamefont {Lake}}]{Anand2017}%
	\BibitemOpen
	\bibfield  {author} {\bibinfo {author} {\bibfnamefont {V.~K.}\ \bibnamefont
			{Anand}}, \bibinfo {author} {\bibfnamefont {D.~L.}\ \bibnamefont
			{Abernathy}}, \bibinfo {author} {\bibfnamefont {D.~T.}\ \bibnamefont
			{Adroja}}, \bibinfo {author} {\bibfnamefont {A.~D.}\ \bibnamefont {Hillier}},
		\bibinfo {author} {\bibfnamefont {P.~K.}\ \bibnamefont {Biswas}}, \ and\
		\bibinfo {author} {\bibfnamefont {B.}~\bibnamefont {Lake}},\ }\href {\doibase
		10.1103/PhysRevB.95.224420} {\bibfield  {journal} {\bibinfo  {journal} {Phys.
				Rev. B}\ }\textbf {\bibinfo {volume} {95}},\ \bibinfo {pages} {224420}
		(\bibinfo {year} {2017})}\BibitemShut {NoStop}%
	\bibitem [{\citenamefont {Anand}\ \emph {et~al.}(2015)\citenamefont {Anand},
		\citenamefont {Bera}, \citenamefont {Xu}, \citenamefont {Herrmannsd\"orfer},
		\citenamefont {Ritter},\ and\ \citenamefont {Lake}}]{Anand2015}%
	\BibitemOpen
	\bibfield  {author} {\bibinfo {author} {\bibfnamefont {V.~K.}\ \bibnamefont
			{Anand}}, \bibinfo {author} {\bibfnamefont {A.~K.}\ \bibnamefont {Bera}},
		\bibinfo {author} {\bibfnamefont {J.}~\bibnamefont {Xu}}, \bibinfo {author}
		{\bibfnamefont {T.}~\bibnamefont {Herrmannsd\"orfer}}, \bibinfo {author}
		{\bibfnamefont {C.}~\bibnamefont {Ritter}}, \ and\ \bibinfo {author}
		{\bibfnamefont {B.}~\bibnamefont {Lake}},\ }\href {\doibase
		10.1103/PhysRevB.92.184418} {\bibfield  {journal} {\bibinfo  {journal} {Phys.
				Rev. B}\ }\textbf {\bibinfo {volume} {92}},\ \bibinfo {pages} {184418}
		(\bibinfo {year} {2015})}\BibitemShut {NoStop}%
	\bibitem [{\citenamefont {Opherden}\ \emph {et~al.}(2018)\citenamefont
		{Opherden}, \citenamefont {Bilitewski}, \citenamefont {Hornung},
		\citenamefont {Herrmannsd\"orfer}, \citenamefont {Samartzis}, \citenamefont
		{Islam}, \citenamefont {Anand}, \citenamefont {Lake}, \citenamefont
		{Moessner},\ and\ \citenamefont {Wosnitza}}]{Opherden2018}%
	\BibitemOpen
	\bibfield  {author} {\bibinfo {author} {\bibfnamefont {L.}~\bibnamefont
			{Opherden}}, \bibinfo {author} {\bibfnamefont {T.}~\bibnamefont
			{Bilitewski}}, \bibinfo {author} {\bibfnamefont {J.}~\bibnamefont {Hornung}},
		\bibinfo {author} {\bibfnamefont {T.}~\bibnamefont {Herrmannsd\"orfer}},
		\bibinfo {author} {\bibfnamefont {A.}~\bibnamefont {Samartzis}}, \bibinfo
		{author} {\bibfnamefont {A.~T. M.~N.}\ \bibnamefont {Islam}}, \bibinfo
		{author} {\bibfnamefont {V.~K.}\ \bibnamefont {Anand}}, \bibinfo {author}
		{\bibfnamefont {B.}~\bibnamefont {Lake}}, \bibinfo {author} {\bibfnamefont
			{R.}~\bibnamefont {Moessner}}, \ and\ \bibinfo {author} {\bibfnamefont
			{J.}~\bibnamefont {Wosnitza}},\ }\href {\doibase 10.1103/PhysRevB.98.180403}
	{\bibfield  {journal} {\bibinfo  {journal} {Phys. Rev. B}\ }\textbf {\bibinfo
			{volume} {98}},\ \bibinfo {pages} {180403(R)} (\bibinfo {year}
		{2018})}\BibitemShut {NoStop}%
	\bibitem [{\citenamefont {Ollivier}\ and\ \citenamefont {Mutka}(2011)}]{IN5}%
	\BibitemOpen
	\bibfield  {author} {\bibinfo {author} {\bibfnamefont {J.}~\bibnamefont
			{Ollivier}}\ and\ \bibinfo {author} {\bibfnamefont {H.}~\bibnamefont
			{Mutka}},\ }\href {\doibase 10.1143/JPSJS.80SB.SB003} {\bibfield  {journal}
		{\bibinfo  {journal} {Journal of the Physical Society of Japan}\ }\textbf
		{\bibinfo {volume} {80}},\ \bibinfo {pages} {SB003} (\bibinfo {year}
		{2011})}\BibitemShut {NoStop}%
	\bibitem [{\citenamefont {Ewings}\ \emph {et~al.}(2016)\citenamefont {Ewings},
		\citenamefont {Buts}, \citenamefont {Le}, \citenamefont {{van Duijn}},
		\citenamefont {Bustinduy},\ and\ \citenamefont {Perring}}]{Horace}%
	\BibitemOpen
	\bibfield  {author} {\bibinfo {author} {\bibfnamefont {R.}~\bibnamefont
			{Ewings}}, \bibinfo {author} {\bibfnamefont {A.}~\bibnamefont {Buts}},
		\bibinfo {author} {\bibfnamefont {M.}~\bibnamefont {Le}}, \bibinfo {author}
		{\bibfnamefont {J.}~\bibnamefont {{van Duijn}}}, \bibinfo {author}
		{\bibfnamefont {I.}~\bibnamefont {Bustinduy}}, \ and\ \bibinfo {author}
		{\bibfnamefont {T.}~\bibnamefont {Perring}},\ }\href {\doibase
		https://doi.org/10.1016/j.nima.2016.07.036} {\bibfield  {journal} {\bibinfo
			{journal} {Nuclear Instruments and Methods in Physics Research Section A:
				Accelerators, Spectrometers, Detectors and Associated Equipment}\ }\textbf
		{\bibinfo {volume} {834}},\ \bibinfo {pages} {132} (\bibinfo {year}
		{2016})}\BibitemShut {NoStop}%
	\bibitem [{\citenamefont {et~al.}(2015)}]{DNS}%
	\BibitemOpen
	\bibfield  {author} {\bibinfo {author} {\bibfnamefont {H.~M.-L.~Z.}\
			\bibnamefont {et~al.}},\ }\href {\doibase
		http://dx.doi.org/10.17815/jlsrf-1-33} {\bibfield  {journal} {\bibinfo
			{journal} {Journal of large-scale research facilities}\ }\textbf {\bibinfo
			{volume} {A27}},\ \bibinfo {pages} {1} (\bibinfo {year} {2015})}\BibitemShut
	{NoStop}%
	\bibitem [{\citenamefont {Huang}\ \emph {et~al.}(2014)\citenamefont {Huang},
		\citenamefont {Chen},\ and\ \citenamefont {Hermele}}]{Huang2014}%
	\BibitemOpen
	\bibfield  {author} {\bibinfo {author} {\bibfnamefont {Y.-P.}\ \bibnamefont
			{Huang}}, \bibinfo {author} {\bibfnamefont {G.}~\bibnamefont {Chen}}, \ and\
		\bibinfo {author} {\bibfnamefont {M.}~\bibnamefont {Hermele}},\ }\href
	{\doibase 10.1103/PhysRevLett.112.167203} {\bibfield  {journal} {\bibinfo
			{journal} {Phys. Rev. Lett.}\ }\textbf {\bibinfo {volume} {112}},\ \bibinfo
		{pages} {167203} (\bibinfo {year} {2014})}\BibitemShut {NoStop}%
	\bibitem [{\citenamefont {Ross}\ \emph {et~al.}(2011)\citenamefont {Ross},
		\citenamefont {Savary}, \citenamefont {Gaulin},\ and\ \citenamefont
		{Balents}}]{Ross2011}%
	\BibitemOpen
	\bibfield  {author} {\bibinfo {author} {\bibfnamefont {K.~A.}\ \bibnamefont
			{Ross}}, \bibinfo {author} {\bibfnamefont {L.}~\bibnamefont {Savary}},
		\bibinfo {author} {\bibfnamefont {B.~D.}\ \bibnamefont {Gaulin}}, \ and\
		\bibinfo {author} {\bibfnamefont {L.}~\bibnamefont {Balents}},\ }\href
	{\doibase 10.1103/PhysRevX.1.021002} {\bibfield  {journal} {\bibinfo
			{journal} {Phys. Rev. X}\ }\textbf {\bibinfo {volume} {1}},\ \bibinfo {pages}
		{021002} (\bibinfo {year} {2011})}\BibitemShut {NoStop}%
	\bibitem [{\citenamefont {Toth}\ and\ \citenamefont {Lake}(2015)}]{TothPRX}%
	\BibitemOpen
	\bibfield  {author} {\bibinfo {author} {\bibfnamefont {S.}~\bibnamefont
			{Toth}}\ and\ \bibinfo {author} {\bibfnamefont {B.}~\bibnamefont {Lake}},\
	}\href {\doibase 10.1088/0953-8984/27/16/166002} {\bibfield  {journal}
		{\bibinfo  {journal} {Journal of Physics: Condensed Matter.}\ }\textbf
		{\bibinfo {volume} {27}},\ \bibinfo {pages} {166002} (\bibinfo {year}
		{2015})}\BibitemShut {NoStop}%
	\bibitem [{\citenamefont {Xu}\ \emph {et~al.}(2020)\citenamefont {Xu},
		\citenamefont {Benton}, \citenamefont {Islam}, \citenamefont {Guidi},
		\citenamefont {Ehlers},\ and\ \citenamefont {Lake}}]{Xu2020}%
	\BibitemOpen
	\bibfield  {author} {\bibinfo {author} {\bibfnamefont {J.}~\bibnamefont
			{Xu}}, \bibinfo {author} {\bibfnamefont {O.}~\bibnamefont {Benton}}, \bibinfo
		{author} {\bibfnamefont {A.~T. M.~N.}\ \bibnamefont {Islam}}, \bibinfo
		{author} {\bibfnamefont {T.}~\bibnamefont {Guidi}}, \bibinfo {author}
		{\bibfnamefont {G.}~\bibnamefont {Ehlers}}, \ and\ \bibinfo {author}
		{\bibfnamefont {B.}~\bibnamefont {Lake}},\ }\href {\doibase
		10.1103/PhysRevLett.124.097203} {\bibfield  {journal} {\bibinfo  {journal}
			{Phys. Rev. Lett.}\ }\textbf {\bibinfo {volume} {124}},\ \bibinfo {pages}
		{097203} (\bibinfo {year} {2020})}\BibitemShut {NoStop}%
	\bibitem [{\citenamefont {L\'eger}\ \emph {et~al.}(2021)\citenamefont
		{L\'eger}, \citenamefont {Lhotel}, \citenamefont {Ciomaga~Hatnean},
		\citenamefont {Ollivier}, \citenamefont {Wildes}, \citenamefont {Raymond},
		\citenamefont {Ressouche}, \citenamefont {Balakrishnan},\ and\ \citenamefont
		{Petit}}]{Leger2021}%
	\BibitemOpen
	\bibfield  {author} {\bibinfo {author} {\bibfnamefont {M.}~\bibnamefont
			{L\'eger}}, \bibinfo {author} {\bibfnamefont {E.}~\bibnamefont {Lhotel}},
		\bibinfo {author} {\bibfnamefont {M.}~\bibnamefont {Ciomaga~Hatnean}},
		\bibinfo {author} {\bibfnamefont {J.}~\bibnamefont {Ollivier}}, \bibinfo
		{author} {\bibfnamefont {A.~R.}\ \bibnamefont {Wildes}}, \bibinfo {author}
		{\bibfnamefont {S.}~\bibnamefont {Raymond}}, \bibinfo {author} {\bibfnamefont
			{E.}~\bibnamefont {Ressouche}}, \bibinfo {author} {\bibfnamefont
			{G.}~\bibnamefont {Balakrishnan}}, \ and\ \bibinfo {author} {\bibfnamefont
			{S.}~\bibnamefont {Petit}},\ }\href {\doibase 10.1103/PhysRevLett.126.247201}
	{\bibfield  {journal} {\bibinfo  {journal} {Phys. Rev. Lett.}\ }\textbf
		{\bibinfo {volume} {126}},\ \bibinfo {pages} {247201} (\bibinfo {year}
		{2021})}\BibitemShut {NoStop}%
	\bibitem [{\citenamefont {Samartzis}(2020)}]{ThesisSamartzis}%
	\BibitemOpen
	\bibfield  {author} {\bibinfo {author} {\bibfnamefont {A.}~\bibnamefont
			{Samartzis}},\ }\emph {\bibinfo {title} {Magnetic investigation of effective
			spin-1/2 magnetic insulators based on 3d and 4f magnetic ions}},\ \href
	{\doibase 10.14279/depositonce-10616} {\bibinfo {type} {Doctoral thesis}},\
	\bibinfo  {school} {Technische Universität Berlin}, \bibinfo {address}
	{Berlin} (\bibinfo {year} {2020})\BibitemShut {NoStop}%
	\bibitem [{\citenamefont {Xu}(2017)}]{ThesisXu}%
	\BibitemOpen
	\bibfield  {author} {\bibinfo {author} {\bibfnamefont {J.}~\bibnamefont
			{Xu}},\ }\emph {\bibinfo {title} {Magnetic properties of rare earth zirconate
			pyrochlores}},\ \href {\doibase 10.14279/depositonce-6408} {\bibinfo {type}
		{Doctoral thesis}},\ \bibinfo  {school} {Technische Universität Berlin},
	\bibinfo {address} {Berlin} (\bibinfo {year} {2017})\BibitemShut {NoStop}%
\end{thebibliography}

%

\end{document}